\documentclass[acmlarge,screen]{acmart}

\usepackage{booktabs} 
\usepackage{algorithm}
\usepackage{algpseudocode}
\usepackage{amsmath, amssymb, amsfonts, amsthm, bm, mathtools}
\usepackage[dvipdfmx]{}

\setcopyright{none}
\begin{document}

\title{Token-Curated Registry with Citation Graph}

\author{Kensuke Ito}
\affiliation{%
  \institution{The University of Tokyo}
  \streetaddress{7-3-1, Hongo, Bunkyo-ward}
  \city{Tokyo}
  \state{Japan}
  \postcode{113-0033}
}
\email{k-ito@g.ecc.u-tokyo.ac.jp}

\author{Hideyuki Tanaka}
\affiliation{%
  \institution{The University of Tokyo}
  \streetaddress{7-3-1, Hongo, Bunkyo-ward}
  \city{Tokyo}
  \state{Japan}
  \postcode{113-0033}
}
\email{tanaka@iii.u-tokyo.ac.jp}


\begin{abstract}
  In this study, we aim to incorporate the expertise of anonymous curators into a token-curated registry (TCR), a decentralized recommender system for collecting a list of high-quality content.
  This registry is important, because previous studies on TCRs have not specifically focused on technical content, such as academic papers and patents, whose effective curation requires expertise in relevant fields.
  To measure expertise, curation in our model focuses on both the content and its citation relationships, for which curator assignment uses the Personalized PageRank (PPR) algorithm while reward computation uses a multi-task peer-prediction mechanism.
  Our proposed {\em CitedTCR} bridges the literature on network-based and token-based recommender systems and contributes to the autonomous development of an evolving citation graph for high-quality content.
  Moreover, we experimentally confirm the incentive for registration and curation in {\em CitedTCR} using the simplification of a one-to-one correspondence between users and content (nodes).
\end{abstract}



\maketitle

\section{Introduction} \label{intro}
For many blockchain-based decentralized applications (DApps), one of the challenges is the reliability of information originating from an {\em off-chain} environment.
This is because the Bitcoin protocol \cite{nakamoto2008bitcoin}, which is the origin of DApps and has a novelty of building a reliable consensus among anonymous users (on a public peer-to-peer network), only computes information generated from an {\em on-chain} environment (i.e., transaction records of Bitcoin).
For example, consider the case of a simple DApp that provides alerts when it rains in a given location.
In this case, while the DApp can ensure the on-chain state transition leads to an alert, it cannot ensure the off-chain fact (used as the trigger) that it has actually rained at the location.
Therefore, most DApps rely on trusted third parties, such as the National Weather Service, for their input\footnote{This situation is often referred to as {\em Oracle problem}.}.
This is in contrast to the Bitcoin protocol, which functions even if the operators of each node are unknown.
Consequently, DApps require an additional protocol in which anonymous users can build reliable consensus on off-chain information to maintain the novelty of the Bitcoin protocol.

A token-curated registry \cite{goldin2017token, goldin2017token2} (TCR) is a DApp for establishing such a protocol. It specializes in compiling a high-quality, reliable list of off-chain content (e.g., restaurants, universities, and webpages) as a recommender system\footnote{See Section \ref{related} for some application examples.}.
Although there are different design patterns among existing TCRs \cite{lockyer2018token}, generally their consensus building is based on a token-staking scheme in which all users can stake their tokens on a binary choice \{accept, reject\} as curators whenever an applicant posts new content to the list\footnote{Wang and Krishnamachari \cite{wang2018enhancing} referred to the TCR for binary choices as {\em objective TCR}.}.
Consensus is the selection that obtains more tokens compared to another selection after a certain period. Moreover, all staked tokens are redistributed among curators who stake their tokens on the consensus side, i.e., token staking intends to yield informative reports from anonymous curators who risk losing their tokens as well as the token price, which is assumed to fluctuate with the quality of the list.
One limitation is that token staking does not reflect expertise in consensus building because, regardless of specialty, any user with certain number of tokens can participate in the curation.
Therefore, the reliability of consensus is restricted under TCRs, which primarily depend only on token staking, particularly when the off-chain content is technical (e.g., academic papers and patents) and requires expertise in specific fields for effective curation.

\begin{figure}[t]
    \centering
    \begin{tabular}{cc}
    \begin{minipage}[t]{0.45\hsize}
    \centering
    \includegraphics[bb=0 0 527 682, width=0.9\hsize]{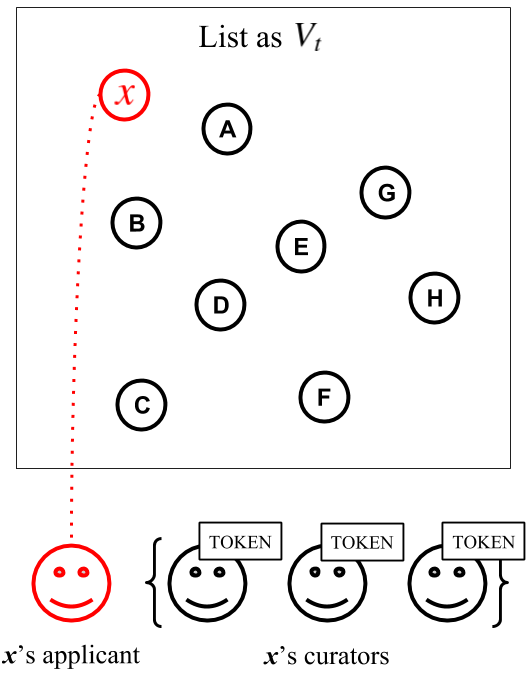} \\
    (a) Existing TCRs
    \end{minipage} &
    \begin{minipage}[t]{0.45\hsize}
    \begin{center}
    \includegraphics[bb=0 0 527 682, width=0.9\hsize]{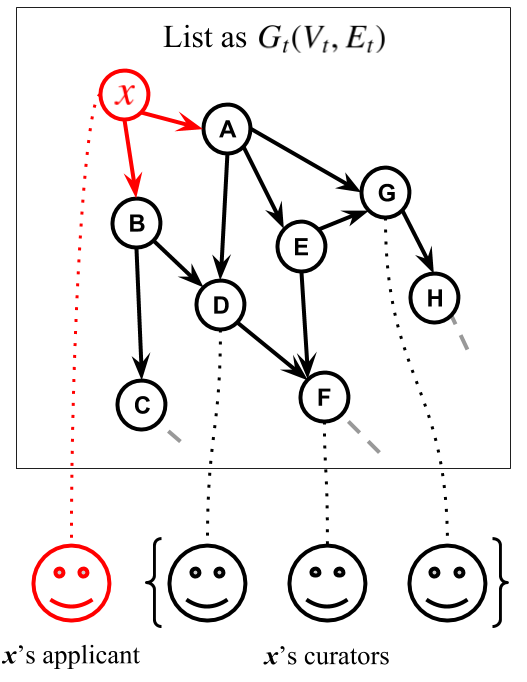} \\
    (b) CitedTCR
    \end{center}
    \end{minipage}
    \end{tabular}
    \caption{TCRs in both cases select curators to decide whether to accept a newly proposed content $x$ into the list of off-chain content $\{A, B, \cdots\}$. However, while existing TCRs (a) manage an unstructured list $V_t$ that can be curated by any user who stakes certain number of tokens, {\em CitedTCR} (b) manages an evolving directed acyclic graph (DAG)-structured list $G_t(V_t,E_t)$ whose curators are assigned according to citation relationships.}
    \label{citedTCR}
\end{figure}

Accordingly, in this study, we aim to incorporate the expertise of anonymous curators into TCRs using a protocol called {\em CitedTCR}, which leverages a citation graph for curator assignment and uses a {\em peer-prediction} mechanism to compute the number of reward tokens paid to the curators.
Fig. \ref{citedTCR} illustrates the role of the citation graph in our protocol.
Fig. \ref{citedTCR} (a) shows that existing TCRs manage an evolving unstructured list (as a set) $V_t$, in which an applicant posts new content (as an element) $x$ and any user can be the curator of $x$ because of token staking.
However, Fig. \ref{citedTCR} (b) shows that {\em CitedTCR} manages an evolving list $G_t(V_t,E_t)$ with a citation graph (i.e., a DAG) structure, in which an applicant posts $x$ and its out-edges $(x, A), (x, B)$ point to existing nodes $\{A, B\}$ as references.
Moreover, curators are stochastically assigned to a given number of users who have posted nodes (e.g., $\{D, F, G\}$) that have both high similarity with $x$'s reference nodes $\{A, B\}$ and high centrality in $G_t$\footnote{As described later, {\em CitedTCR} uses the Personalized PageRank (PPR) algorithm to measure both similarity and centrality.}.
{\em CitedTCR} assigns appropriate curators in a manner similar to the academic peer-review process, in which researchers who have produced high-quality papers with a large number of citations are more likely to be selected as reviewers in their field of expertise.
Note that this form of curator assignment serves as an incentive for applicants to register high-quality content in {\em CitedTCR} because users may have more opportunities to obtain reward tokens as curators if their content in $G_t$ attracts a large number of citations\footnote{We will confirm the strength of this incentive in Section \ref{experiments}.}\footnote{As a similar incentive, TCRs using the token staking often require applicants to stake a certain amount of their token on $\{$accept$\}$ choice.}.
The citation graph serves as a proxy for the expertise of anonymous curators; therefore, the reliability of $G_t$ from the perspective of both curation and registration is ensured.

Peer prediction is a mechanism of game theory for eliciting informative reports for tasks with no ground truth, such as the peer review of academic papers and online product reviews by consumers.
In particular, peer prediction compares user reports for the same task to create a {\em truthful} (known as {\em strategy-proof} or {\em incentive-compatible}) environment, in which no user can obtain a higher utility by any possible strategy deviating from the user's true preferences \cite{nisan2007algorithmic}.
{\em CitedTCR} uses peer prediction for reward computation, in which it is assumed that the assigned curators can obtain newly issued reward tokens if they return a binary signal $\{$accept, reject$\}$ as a report for $x$ and $x$'s citation relationships\footnote{As described in Section \ref{model}, $x$ is listed including its citation relationship if the number of \{accept\} reports exceeds a given threshold.}.
This mechanism addresses two problems in the token-staking scheme, which is even more critical under {\em CitedTCR}.
The first problem is the risk of strategic misreports (such as collusion) among curators.
Although this has been discussed for existing TCRs \cite{falk2018token, curatethis}, token staking becomes more vulnerable to this risk in {\em CitedTCR} because {\em CitedTCR} assigns a fixed number of homogeneous curators with similar expertise.
The second problem is the lack of incentive to participate in consensus building because of the risk of losing staked tokens (see Appendix A). Note that strengthening the weak incentive of token staking is a common topic in TCRs \cite{wang2018enhancing}.
Stronger incentives are particularly important for {\em CitedTCR} in which reports elicited from assigned curators are the key for reflecting expertise in $G_t$.
Therefore, rather than token staking, we use a peer-prediction mechanism that provides maximum (new) rewards for informative reports.

{\em CitedTCR} is thus a hybrid of token-based and network-based recommender systems because it recommends both $V_t(G_t)$ curated by tokens and curators assigned according to $G_t$.
In this study, as a first step of this hybrid approach, we used the Personalized PageRank \cite{haveliwala2002topic} (PPR) algorithm for curator assignment, and a peer prediction mechanism called DG13 proposed by Dasgupta and Ghosh \cite{dasgupta2013crowdsourced} for reward computation.
In addition to their popularity, both PPR and DG13 have several favorable properties for {\em CitedTCR} as demonstrated in Sections \ref{related} and \ref{model}.
Moreover, we assume that users in this study have a one-to-one correspondence with $V_t(G_t)$.
This assumption is intended to simplify the curation process into a state transition in $G_t$; its details are discussed in Section \ref{model}.

The remainder of this paper is organized as follows.
In Section \ref{related}, we introduce related studies and contributions from the perspective of three components: TCR, the PPR algorithm, and a peer-prediction mechanism.
In Section \ref{model}, we describe the specification of {\em CitedTCR}, including the role of PPR and DG13.
In Section \ref{experiments}, we examine the practical utility of our proposal using two step-wise simulations with the citation graph of academic papers.
Finally, in Section \ref{conclusion}, we concludes the paper with a summary of achievements and remaining concerns.

\section{Related Work} \label{related}

\subsection{TCR}
Since Goldin \cite{goldin2017token, goldin2017token2} proposed the initial design in 2017, TCRs have been implemented in a number of applications such as the adChain registry\footnote{\url{https://metax.io/en/products/adchain_registry/}, (accessed April 20, 2019)} for webpages, the Ocean protocol\footnote{\url{https://oceanprotocol.com/}, (accessed April 20, 2019)} for user reputations, and the Civil registry\footnote{\url{https://registry.civil.co/registry/approved}, (accessed April 20, 2019)} for news articles.
Because TCR is a recent development, most discussion at present focus on blog articles whose topics vary from the classification of design patterns \cite{lockyer2018token} to critical examinations of token staking \cite{curatethis, designflaws}.
A reading list curated by the blockchain community \cite{tcrlist, tcrlist2} would be helpful for summarizing this discussion.
In addition to blog articles, TCRs have been examined in academic papers, primarily from a game-theoretic perspective.
For example, Asgaonkar and Krishnamachari \cite{asgaonkar2018token} presented a mathematical foundation of the TCR 1.1 model \cite{goldin2017token2} to determine the sufficient conditions for each consensus at equilibrium.
Wang and Krishnamachari \cite{wang2018enhancing} introduced enhanced token staking with a new issuance of reward tokens to create an incentive to participate in consensus building.
Moreover, Falk and Tsoukalas \cite{falk2018token} used an axiomatic approach to demonstrate the limitations of a token-staking scheme, in which the expected rewards are proportional to the amount of staking.

As mentioned in Section \ref{intro}, in this study, we aim to incorporate the expertise of anonymous curators into TCRs using a combination of citation graphs and peer prediction (i.e., PPR and DG13).
This approach is novel because previous studies and blog articles on TCRs have not explicitly addressed the mechanism for technical content, such as academic papers and patents, whose effective curation requires expertise in relevant fields.

\subsection{The PPR algorithm}
The PPR \cite{haveliwala2002topic} algorithm, originally named topic-sensitive PageRank, is an extension of the PageRank \cite{brin1998anatomy, page1999pagerank} algorithm and computes a score of importance for each node from the viewpoint of the entire network structure.
While the PageRank score originates from a random walk on the network, PPR allows this random walk to return to the predetermined set of nodes with a given probability\footnote{For this property, PPR is often referred to as the random walk with restart (RWR) algorithm.}, thereby adapting the score to recommender systems (see Section \ref{ppr} for details).
In many recommender systems using PPR, {\em CitedTCR} is most closely related to the PaperRank algorithm proposed by Gori and Pucci \cite{gori2006research}, which applies PPR to a citation graph of academic papers to generate useful paper-to-paper recommendations.
Moreover, PPR is a component of several paper-to-reviewer assignment systems \cite{liu2014robust, kuccuktuncc2012recommendation} that attempt to recommend appropriate peer reviewers for a submitted paper.

From the perspective of PPR, this study provides contributions such as {\em CitedTCR} bridging the literature on network-based and token-based recommender systems for the first time to strengthen the reliability of the consensus.
New economy movement (NEM) \cite{nem2018nem} is a representative precedent of blockchain-based protocols that leverage a network structure for consensus building.
However, NEM is not specific to TCRs and manages on-chain transaction records using a network-based score different from that of PPR.

\subsection{Peer-prediction mechanism}
Peer prediction was first introduced by Miller et al. \cite{miller2005eliciting} as an application of the proper scoring rule \cite{gneiting2007strictly} and game theory\footnote{See the textbook \cite{faltings2017game}, for more comprehensive review on peer-prediction method and other information elicitation models.}.
To model the problem of eliciting private information, reward (score) computation assumes an environment in which each user reports probabilistic but correlated signals based on the assigned tasks.
As examined by Jurca and Faltings \cite{jurca2005enforcing}, a common problem in the mechanism proposed by Miller et al. and subsequent mechanisms is that the computation has multiple Nash equilibria, including uninformative ones in which elicited reports are independent of the true signals\footnote{Uninformative equilibria are designated as a {\em blind agreements} in original DG13 \cite{dasgupta2013crowdsourced}.}; e.g., the same signals or random signals are always reported to avoid the effort of observation.
As a solution to this problem, Dasgupta and Ghosh \cite{dasgupta2013crowdsourced} proposed a multi-task peer-prediction mechanism called DG13 that assigns multiple tasks to one user and computes rewards for one task using the reports produced for other tasks.
Under the assumption of positively correlated binary signals, DG13 ensures strong truthfulness \cite{shnayder2016informed}, in which an equilibrium by informative reports has the highest rewards among other realistic equilibria (see Section \ref{dg13explain} for details).
{\em CitedTCR} uses DG13 because the abovementioned properties of multi-tasking, strong truthfulness, and binary signals are compatible with the general settings of TCRs, in which curators evaluate multiple content using binary choices.

To our knowledge, {\em CitedTCR} is the first proposal that uses a peer-prediction mechanism in TCRs.
This proposal presents an approach that can overcome the aforementioned two problems in the token staking.
In addition to DG13, recent studies on peer prediction have discussed topics relevant to TCRs.
For example, Agarwal et al. \cite{agarwal2017peer} proposed a multi-task mechanism that assigns appropriate tasks to heterogeneous users (with various propensities) based on accumulated reports.
This can contribute to TCRs with expertise as an approach different from citation graphs.
Goel et al. \cite{goel2019decentralized} assessed the robustness of a peer-prediction mechanism for the case in which an incentive for misreporting exists outside the system with an application to decentralized oracles\footnote{Decentralized oracle is a broader concept than TCR, which includes every DApp responsible for consensus-building on off-chain contents, i.e., TCR can be interpreted as one of the decentralized oracle systems. The term decentralized oracle is often used in the context of prediction market, and representative platforms (e.g., Augur \cite{peterson2015augur}, Gnosis \cite{team2017gnosis}) use the token staking for their consensus building as with the case of TCRs.}.
Their assessment can be applicable to TCRs with a design similar to that of decentralized oracles.

\section{Model} \label{model}
In this section, we describe the specification of {\em CitedTCR} as a state transition closed on list $G_{t}$.
This simplification, achieved by several assumptions, including the aforementioned one-to-one correspondence, is useful for an algorithmic expression and for the experimental simulations described in Section \ref{experiments}.
Moreover, we present details of PPR and DG13 that clarify how these components contribute to curation in {\em CitedTCR}.

\subsection{Setup}
As depicted in Fig. \ref{citedTCR} (b), our protocol deals with an evolving DAG-structured list $G_t(V_t, E_t)$, where $V_t$ denotes the set of registered content and $E_{t}\subseteq V_{t}\times V_{t}$ denotes their citation relationships.
Although $G_t$ is managed by a set of users $U_t$ (as with other DApps), we impose the following assumption on the management of $G_t$.

\vspace*{12pt}
\noindent
{\bf Assumption~1} One-to-one correspondence: Suppose that there is a one-to-one correspondence between $U_t$ and $V_t$, i.e., $f: U_t\to V_t$ is bijective.

\vspace*{12pt}
\noindent
A one-to-one correspondence indicates an environment in which a user can neither post more than one content nor share one content as a co-applicant.
This setting frees our model from several complex problems in DApps, such as spamming and sybil attacks, and makes curator assignment equivalent to node selection in $G_{t}$.

We further assume that only one node $x$ proposes an additional citation graph $\dot{G_{t}}$ (composed of the references of $x$ and $x$) in each period, and $\dot{G_{t}}$ is not delisted once it is accepted into $G_{t}$.
This assumption and one-to-one correspondence make it possible to represent {\em CitedTCR} as a state transition $\{G_{t}\}_{t=0}^{\infty}$ that repeatedly determines whether to accept $\dot{G_{t}}$ in each period\footnote{Therefore, when managing {\em CitedTCR}, we need to prepare in advance an initial state $G_{0}$ with a sufficient number of nodes and edges.}.
In particular, the transition from $G_{t}$ to $G_{t+1}$ can be summarized as follows:

\noindent
\begin{enumerate}
\item A new node $x$ proposes $\dot{G_{t}}(\{x\} \cup V_{x}, E_{x})$ to $G_t(V_t, E_t)$, where $V_{x}$ denotes the set of $x$'s reference nodes (i.e., $V_{x}\subseteq V_{t}$), and $E_{x}$ denotes directed edges from $x$ to $V_{x}$.
\item Curator assignment: Select $n (\geq 2)$\footnote{The condition $n \geq 2$ is important for DG13 mechanism as we will see in Section \ref{dg13explain}.} of nodes $\dot{C_{t}} = \{1,2,\cdots,n\}$ as curators from $V_{t} \setminus V_{x}$, where $n$ is an exogenous variable\footnote{Thus, $|V_{x}|$ needs an upper limit number which must satisfy $|V_{t}\setminus V_{x}|\geq n$ for all $t$.}.
\item Collect $n$ reports $\dot{R_{t}}=\{r_{1}^{\dot{G_{t}}},r_{2}^{\dot{G_{t}}}, \cdots, r_{n}^{\dot{G_{t}}} \}$ from $\dot{C_{t}}$, where $r_{c}^{\dot{G_{t}}}\in \{0, 1\}$ denotes curator $c$'s report for $\dot{G_{t}}$. Here, $r=0$ and $r=1$ designate {\em reject} and {\em accept}, respectively.
\item Reward computation: Compute rewards $\Theta=\{\theta_1^{\dot{G_{t}}}, \theta_2^{\dot{G_{t}}}, \cdots, \theta_n^{\dot{G_{t}}}\}$ for $\dot{C_{t}}$.
\item Update $G_{t}$ to $G_{t+1}$. $G_{t+1}$ includes $\dot{G_{t}}$ only if $\dot{R_{t}}$ has $m (\leq n)$ or more number of $r=1$, where $m$ is another exogenous variable.
\end{enumerate}

A pseudocode can be used to convert this state transition into Algorithms $1$ and $2$, in which, as commented, curator assignment (step 2) uses PPR, and reward computation (step 4) uses DG13.
These algorithms include the following two properties.
First, they integrate steps 2 and 3 as the C\textsc{uration}$(n,C,R,G)$ function (Algorithm $2$), which returns a set of reports $R$ for the following four arguments: $n$, the number of reports; $C$, the set of nodes that are candidates for the curator; $R$, the initial value of the set of reports; and $G$, the graph containing $C$.
This integration is intended to handle a case in which assigned curators do not provide their reports within a given period of time.
In this case, C\textsc{uration}$(n,C,R,G)$ continues to reselect new nodes as replacements for unresponsive curators until it collects $n$ reports.
Second, they return not only $G_{t+1}$ and $\Theta$ but also the stock of reports $R_{t+1}$.
This property is specific to DG13, whose reward computation leverages both the flow $\dot{R_{t}}$ and stock $R_{t}$ of elicited reports as one of the multi-task peer prediction mechanisms.
Algorithm $1$ can be simplified by adopting other intratemporal mechanisms such as token staking.

\begin{algorithm}[t]
\caption{State transition in {\em CitedTCR}}
\begin{algorithmic}[1]
    \State{$G_{t}(V_{t}, E_{t}) \gets$ list in period $t$}
    \State{$\dot{G_{t}}(\{x\} \cup V_{x}, E_{x}) \gets$ proposal by $x$}
    \State{$\{m,n\} \gets$ exogenous variables}
    \State{$R_{t} \gets$ stock of reports until period $t$} \Comment{Specific to DG13}
    \State{$\dot{R_{t}} \gets$ \Call{Curation}{$n, V_{t} \setminus V_{x}, \{\emptyset\}, G_{t}$}} \Comment{See Algorithm 2}
    \State{Compute rewards $\Theta$ with $R_t$ and $\dot{R_{t}}$} \Comment{Use DG13}
    \State{\Return $\Theta$}
    \State{$R_{t+1}\gets R_{t} \cup \dot{R_{t}}$} \Comment{Specific to DG13}
    \State{\Return $R_{t+1}$} \Comment{Specific to DG13}
    \If{$m \geq |\{r\in \dot{R_{t}}|r=1\}|$}
    \State{$G_{t+1}\gets G_{t}$}
    \Else
    \State{$G_{t+1}\gets G_{t} \cup G_{x}$}
    \EndIf
    \State{\Return $G_{t+1}$}
\end{algorithmic}
\end{algorithm}

\begin{algorithm}[t]
\caption{Report collection and curator assignment in {\em CitedTCR}}
\begin{algorithmic}[1]
    \Function{Curation}{$n, C, R, G$}
    \State{$C^{\prime}\gets$ $n$ curators selected from $C$ in $G$} \Comment{Use PPR}
    \State{$R^{\prime}\gets$ reports collected from $C^{\prime}$ within a given period of time}
    \State{$R\gets R \cup R^{\prime}$}
    \If{$|R^{\prime}|=n$}
    \State{\Return $R$}
    \Else
    \State{$n\gets n - |R^{\prime}|$}
    \State{$C\gets C\setminus C^{\prime}$}
    \State{\Call{Curation}{$n, C, R, G$}}
    \EndIf
    \EndFunction
\end{algorithmic}
\end{algorithm}

\subsection{PPR for curator assignment} \label{ppr}
PPR is an algorithm that recommends relatively important nodes to a given node through iterative random walking on a network (as a Markov chain).
{\em CitedTCR} uses PPR for a curator assignment that selects $\dot{C_{t}}$ as important nodes for $x$.
In the example presented in Fig. \ref{citedTCR} (b), the set of curators $\dot{C_{t}}=\{D,F,G\}$ is selected from $\{A,B,C,\cdots\} \setminus \{A,B\}$ for the assessment of $\dot{G_t}(\{x,A,B\},\{(x,A),(x,B)\})$, where $\dot{C_{t}}$ is regarded as important nodes from the standpoint of $x$ with the reference $V_{x}=\{A,B\}$.
Nodes such as $V_{x}$ are often referred to as {\em base nodes} in the PPR context, and are the key for computing relative importance.

To quantify the process of random walking, PPR leverages a transition matrix $\bm{P}$, which in our case is $|V_{t}|\times |V_{t}|$, and an element $p_{ij}$ designates the probability of transition from node $i$ to node $j$.
In the random walk as a Markov chain, the value of $p_{ij}$ becomes the reciprocal of node $i$'s out-degree.

The simplified PageRank\footnote{Although original paper \cite{page1999pagerank} uses Simplified PageRank as an introduction of model description, PageRank is the dominant eigenvector of the matrix $\bm{P}_{PR} = (1-\alpha)\bm{P} + \alpha (1/|V_{t}|)\bm{1}$, where $\bm{1}$ is $|V_{t}|\times |V_{t}|$ matrix whose elements are all $1$. Namely, $\bm{P}_{PR}$ quantifies the random walking which, with probability $\alpha$, jumps to one of all existing nodes uniformly at random (random-surfer model). This is to make PageRank work even in the directed network including dead-end loop or the node with no out-edges.} score of $V_{t}$ is the dominant eigenvector (for eigenvalue $1$) of $\bm{P}$, which indicates the steady-state probability distribution as a result of iterative random walking.
Moreover, the PPR score of $V_{t}$ is the dominant eigenvector of $\bm{P}_{PPR}$, which has the following modification to $\bm{P}$ \cite{haveliwala2002topic}:

\begin{displaymath}
  \bm{P}_{PPR} = (1-\alpha)\bm{P} + \alpha \frac{1}{|V_{x}|}\bm{B},
\end{displaymath}

\noindent
where $\bm{B}$ is an additional $|V_{t}|\times |V_{t}|$ matrix whose element $b_{ij}$ becomes $1$ if $j$ is included in base nodes $V_{x}$; otherwise, it becomes $0$ (i.e., $b_{iA}$ and $b_{iB}$ become $1$ for all $i$ and other elements become $0$ in Fig. \ref{citedTCR} (b)).
We can interpret $\bm{B}/|V_{x}|$ as another transition matrix in which all nodes in $V_{t}$ must jump to one node selected from $V_{x}$ uniformly at random.
Thus, $\bm{P}_{PPR}$ is the linear combination of the two transition matrices $\bm{P}$ and $\bm{B}/|V_{x}|$ that represents biased random walking, which jumps to one of the base nodes with probability $\alpha$ in each step.
Here, $\alpha \in [0,1]$ is called a damping factor, and it can adjust the strength of bias as an exogenous parameter ($\alpha = 0.15$ in most cases).
{\em CitedTCR} stochastically selects $n$ curators in each period according to the PPR computed from $\bm{P}_{PPR}$.

Below, we discuss three properties in this application of PPR.
First, similar to PaperRank \cite{gori2006research}, {\em CitedTCR} considers $G_{t}$ to be undirected when using PPR.
This is important because if PPR were on a DAG structure, its score would focus on the nodes with no out-edges (i.e., the oldest content in the case of citation graph) and would thus be unreliable for recommender systems.
Second, as already mentioned, {\em CitedTCR} excludes $V_{x}$ from the candidates of $\dot{C_{t}}$.
Although PPR scores high for the base nodes (reference nodes for $x$),
we do not select them to avoid biased curation, in which assigned curators accept $x$ simply to increase their number of citations\footnote{Note that even this modification cannot completely eliminate the biased curation, as long as the curation affects the future structure of $G_{t}$. Analyzing the strength of this bias is one of our future tasks.}.
Third, {\em CitedTCR} can encourage users to register high-quality content in $G_{t}$, even though the frequency with which they become curators is weighted by PPR.
This is experimentally confirmed in Section \ref{experiments} using the PageRank score in $G_{t}$ as a proxy for quality.

\subsection{DG13 for reward computation} \label{dg13explain}
DG13 and other peer-prediction mechanisms aim to elicit truthful information from the environment, in which users report the quality of a task. For example, in {\em CitedTCR}, $n$ assigned curators $\dot{C_{t}} = \{1,2,\cdots,n\}$ provide reports $\dot{R_{t}}=\{r_{1}^{\dot{G_{t}}},r_{2}^{\dot{G_{t}}}, \cdots, r_{n}^{\dot{G_{t}}} \}$ on the quality of $\dot{G_{t}}$.
To confirm whether a report is truthful, peer prediction assumes the stochastic signal $s$, which any $c\in \dot{C_{t}}$ can observe from $\dot{G_{t}}$ and can use as input information for $r_{c}^{\dot{G_{t}}}$.
DG13 focuses on binary signals $s\in \{0,1\}$ and binary reports $r(s)\in \{0,1\}$ ($0$: reject; $1$: accept). We use notation $s_{c}^{\dot{G_{t}}}$ in the same manner as in reporting, i.e., curator $c$ accepts adding $\dot{G_{t}}$ to the $G_{t}$ if $r_{c}^{\dot{G_{t}}}(s_{c}^{\dot{G_{t}}})=1$ and rejects it if $r_{c}^{\dot{G_{t}}}(s_{c}^{\dot{G_{t}}})=0$.
This report is truthful in the $r_{c}^{\dot{G_{t}}}(0)=0$ or $r_{c}^{\dot{G_{t}}}(1)=1$ case and non-truthful in the $r_{c}^{\dot{G_{t}}}(0)=1$ or $r_{c}^{\dot{G_{t}}}(1)=0$ case.
Note that $r_{c}^{\dot{G_{t}}}$ and $s_{c}^{\dot{G_{t}}}$ are sometimes denoted $r_{c}$ and $s_{c}$ when their task does not need to be emphasized.

\begin{figure}[t]
    \centering
    \includegraphics[bb=0 0 1489 654, width=0.7\hsize]{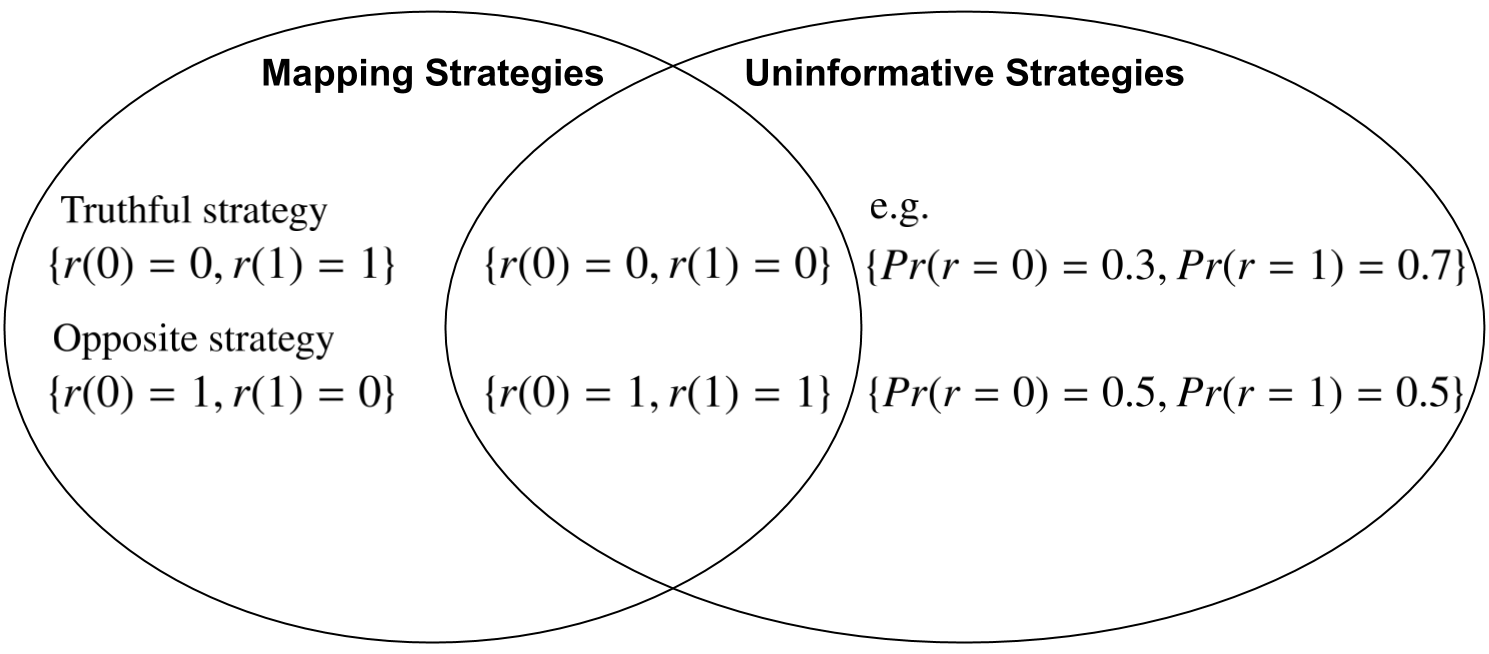}
    \caption{Nodes (curators) can use either mapping or uninformative strategies for reporting. The strategy of always reporting 0 or always reporting 1 can be classified as both mapping and uninformative strategies.}
    \label{venn}
\end{figure}

We add two more assumptions that are common in the literature on peer prediction for binary signals \cite{jurca2005enforcing, witkowski2012peer, dasgupta2013crowdsourced}.
First, $s$, observed by each curator from each task, is positively correlated.
Accordingly, when we randomly select another curator $\hat{c}\in \dot{C_{t}}$, both $Pr(s_{c}=0|s_{\hat{c}}=0) > Pr(s_{c}=0)$ and $Pr(s_{c}=1|s_{\hat{c}}=1) > Pr(s_{c}=1)$ hold for all $c$ and $\hat{c}$, regardless of the tasks\footnote{Accordingly, $Pr(s_{c}=1|s_{\hat{c}}=0) < Pr(s_{c}=1)$ and $Pr(s_{c}=0|s_{\hat{c}}=1) < Pr(s_{c}=0)$ hold, simultaneously.}.
This requires the propensity of assigned $\dot{G_{t}}$ and the peer curators of $c$ to be somewhat homogeneous\footnote{The homogeneity required for positively correlated signals is not as strong in binary signals as in multiple signals.} throughout each period.
{\em CitedTCR} with a citation graph ensures such an environment by curator assignment based on PPR; this is unlike recent multi-task peer prediction \cite{mandal2016peer, agarwal2017peer}, which becomes complex to relax this assumption.
The second assumption is that each curator must select one reporting {\em strategy} from feasible choices.
The set of feasible strategies in our model, presented in Fig. \ref{venn}, is the union of mapping strategies and uninformative signal-independent strategies.
Mapping strategies follow a mapping rule from signals to reports; however, the reports in uninformative strategies follow a given stochastic distribution independent of the observed signals.
For the four possible mapping strategies under the assumption of binary signals, we specifically define a strategy that always reports truth as a {\em truthful} strategy, and a strategy that always reports non-truth as an {\em opposite} strategy.

Finally, if we let $R_{c}\subset R_{t}$ be the set of all (intertemporal) reports that $c$ has provided for multiple $\dot{G_{t}}$s, and let $R_{c}^{*}$ be a special case in which all elements are truthful reports (i.e., $c$ adopts a truthful strategy), the achievement of DG13 can be defined as follows:

\vspace*{12pt}
\noindent
{\bf Definition~1} Strong truthfulness: A mechanism satisfies {\em strong truthfulness} if $\mathbb{E}\left[ \theta_{c}^{\dot{G_{t}}}\mid R_c^{*}, R_{\hat{c}}^{*} \right] \geq \mathbb{E}\left[ \theta_{c}^{\dot{G_{t}}}\mid R_c, R_{\hat{c}}\right]$ holds for all $c, \hat{c}, R_c, R_{\hat{c}}, $ and $\dot{G_{t}}$, where equality occurs only when both $c$ and $\hat{c}$ adopt the opposite strategy\footnote{The original definition \cite{shnayder2016informed} generalizes both truthful strategy and opposite strategy as a {\em permutation strategy} to encompass the case of multiple (non-binary) signals.}.

\vspace*{12pt}
\noindent In other words, compared to any other strategy, the mechanism satisfying strong truthfulness can assign strictly higher expected rewards $\mathbb{E}\left[\theta_{c}^{\dot{G_{t}}}\right]$ to the equilibrium by truthful strategies for almost all cases.

DG13, as a multi-task peer prediction mechanism, computes $c$'s reward $\theta_{c}^{\dot{G_{t}}}$ using not only the reports that $c$ and randomly selected $\hat{c}$ produced in period $t$ (i.e., $r_{c}^{\dot{G_{t}}}$, $r_{\hat{c}}^{\dot{G_{t}}}$) but also all reports that $c$ and $\hat{c}$ produced until period $t$ (i.e., $R_c$, $R_{\hat{c}}$).
According to the original report \cite{dasgupta2013crowdsourced} and a subsequent report for its generalization \cite{shnayder2016informed}, DG13 can be formulated as

\begin{displaymath}
  \theta_{c}^{\dot{G_{t}}} = \delta_{r_{c}^{\dot{G_{t}}},r_{\hat{c}}^{\dot{G_{t}}}} - \delta_{r_{c}\in \{R_{c}\setminus \dot{R_{t}}\},r_{\hat{c}}\in \{R_{\hat{c}}\setminus \dot{R_{t}}\}},
\end{displaymath}

\noindent
where we use the following Kronecker's delta for the sake of convenience:
    \begin{align}
        \delta_{x,y} \ =\
        \begin{cases}
            1 & {\rm if}\;\; x = y \\
            0 & {\rm if}\;\; x \neq y
        \end{cases} \nonumber
    \end{align}

\noindent
Here, $\delta_{r_{c}^{\dot{G_{t}}},r_{\hat{c}}^{\dot{G_{t}}}}$ is the reward for curation in period $t$.
It is apparent that a value of $1$ is obtained when two reports for $\dot{G_{t}}$ return the same signal $(r_{c}^{\dot{G_{t}}}, r_{\hat{c}}^{\dot{G_{t}}})=(0,0)$ or $(1,1)$; otherwise, the value is $0$.
$\delta_{r_{c}\in \{R_{c}\setminus \dot{R_{t}}\},r_{\hat{c}}\in \{R_{\hat{c}}\setminus \dot{R_{t}}\}}$ is a type of penalty that randomly selects two reports $r_{c}$ and $r_{\hat{c}}$ produced by each curator before period $t$ and compares them in the same manner.
Assuming that $c$ and $\hat{c}$ always report $1$ for assigned tasks irrespective of the signals, $\theta_{c}^{t}=0$ holds because the penalty term becomes $1$ even though $r_{c}^{\dot{G_{t}}}$ and $r_{\hat{c}}^{\dot{G_{t}}}$ always represents a reward of $1$.
A similar result would be derived for the case of a 50-50 uninformative strategy (i.e., $Pr(r=0) = Pr(r=1) = 0.5$) because the expected value of reward terms and penalty terms both become $0.5$.
Although $\theta_{c}^{\dot{G_{t}}}$ takes the interval $[-1,1]$ because of the penalty, all rewards can be non-negative by adding $1$ to all $\theta_{c}^{\dot{G_{t}}}$ as a basic reward.

Dasgupta and Ghosh \cite{dasgupta2013crowdsourced} indicated that the expected (net) reward $\mathbb{E}\left[ \theta_{c}^{\dot{G_{t}}} \right]$ is maximized in the equilibrium in which all curators adopt a truthful strategy by exerting efforts on signal observation under the assumption of positively correlated signals.

\vspace*{12pt}
\noindent
{\bf Theorem~1}: DG13 satisfies strong truthfulness.

\vspace*{12pt}
\noindent
See Appendix B for the proof of this theorem.

Note that DG13 in {\em CitedTCR} must collectively compute rewards for previous reports after $c$ and $\hat{c}$ both finish reporting three times.
Three is the number that satisfies the minimum requirements for establishing multi-task peer prediction without loss of generality \cite{shnayder2016informed}: (i) two users, (ii) three total tasks, and  (iii) two or more tasks per user, including at least one common task.
Although each node curates many $\dot{G_{t}}$s during $\{G_{t}\}_{t=0}^{\infty}$ (as long as it has high quality), {\em CitedTCR} with iterative reward computation cannot satisfy (iii) when either $c$ or $\hat{c}$ produces a report for the first time.
Thus, we postpone reward computation until both $c$ and $\hat{c}$ are sure to meet all minimum requirements by three reports\footnote{Two reports cannot satisfy (ii) if $c$ and $\hat{c}$ share two tasks.}; thus, DG13 can elicit truthful reports from curators.

\section{Experimental Studies} \label{experiments}
Although Section \ref{model} describes the utility of PPR and DG13, our study must assess how their combination contributes to the construction of the reliable list $G_{t}$.
In this section, we perform this assessment experimentally using two step-wise simulations that are both based on the DAG-structured dataset formatted from the arXiv high-energy physics theory (HEP-TH) citation network.
In particularly, the simulation first uses only PPR to examine the strength of the incentive for registering high-quality content. It then incorporates DG13 to confirm the incentive for eliciting informative reports.
All materials used for this experiment are available in the Github repository\footnote{\url{https://github.com/knskito/materials_CitedTCR}}.

\begin{figure}[t]

    \centering
    \includegraphics[bb=0 0 4150 3035, width=0.8\hsize]{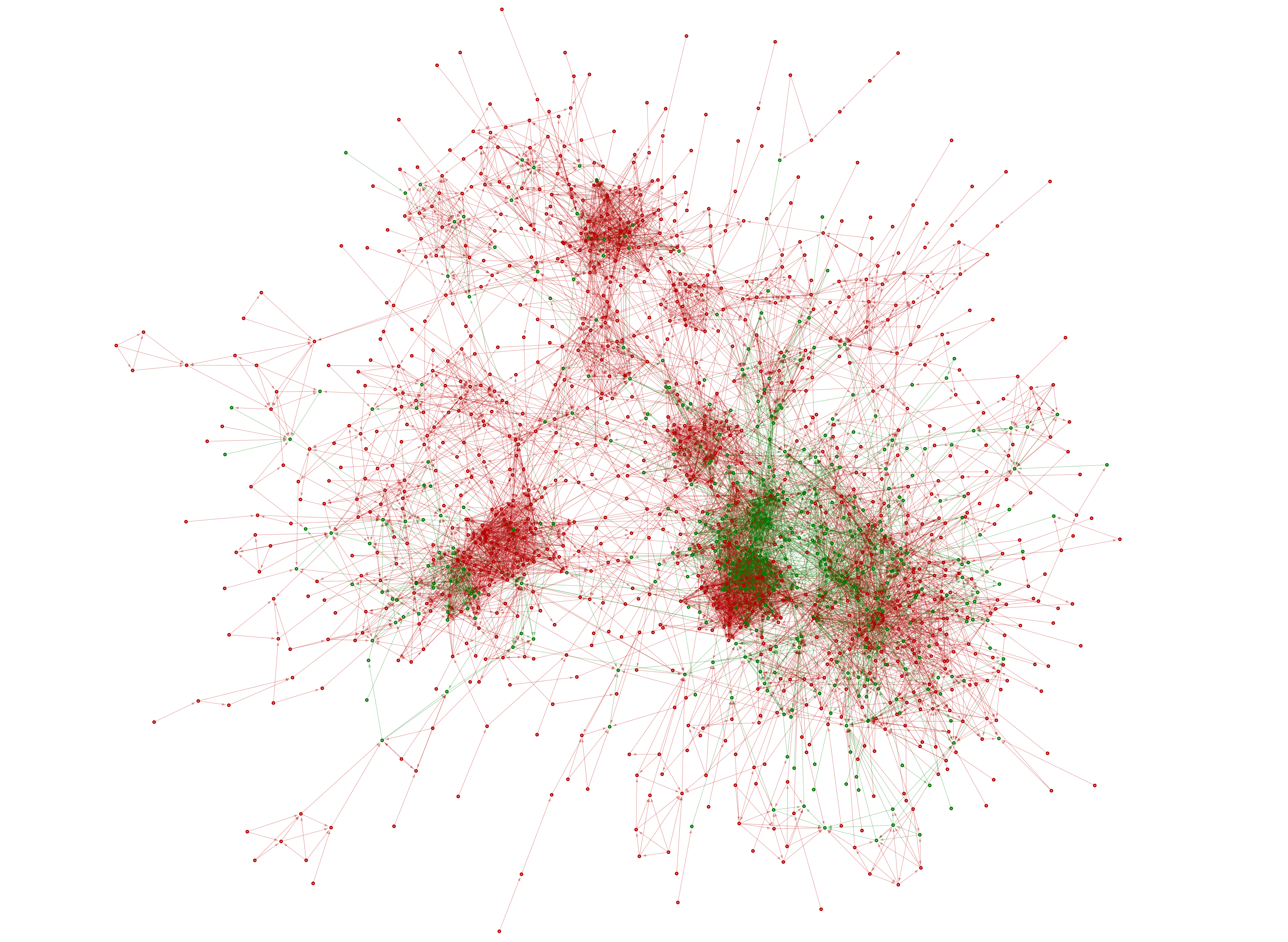}
    \caption{Our experiments use a DAG structure with 1,421 time-ordered nodes, where green represents the citation relationships of the first 421 nodes, while red represents the citation relationships of the last 1,000 nodes. We consider the state transition $\{G_{t}\}_{t=0}^{1000}$ by letting the green (subgraph) be $G_{0}$.}
    \label{network}

\end{figure}

\subsection{Dataset}
The arXiv HEP-TH citation network is a dataset provided by Stanford Network Analysis Project\footnote{\url{https://snap.stanford.edu/data/cit-HepTh.html}}  (SNAP), which contains the citation relationships of academic papers in the HEP-TH category submitted from January 1993 to April 2003.
We selected one component with 1,421 papers since January 2000, and constructed a DAG structure as depicted in Fig. \ref{network} (powered by Cytoscape \cite{shannon2003cytoscape}). Here, the green component represents the citation relationships of the first 421 nodes, while the red component represents the citation relationships of the last 1,000 nodes (i.e., the green part is a subgraph of the DAG structure).
Our experiments consider the green component the initial state $G_{0}$ and consider the state transition $\{G_{t}\}_{t=0}^{1000}$ by sequentially adding the nodes and edges in the red component to $G_{t}$.

\subsection{Incentive for registering high-quality content}
Thus far, we have assumed that {\em CitedTCR} tends to select curators more frequently from nodes that are regarded as important in $G_{t}$, which serves as an incentive for users to register high-quality content.
However, this assumption is not obvious because the curator assignment in each period is weighted by the PPR algorithm, which excludes even base nodes from the candidate list.
To determine the true strength of the incentive for registering high-quality content, our first experiment computes the correlation between the frequency distribution for $1,421$ nodes to be selected as a curator because of sequential assignments up to $G_{1000}$, and the (not simplified) PageRank score for $1,421$ nodes in $G_{1000}$\footnote{We set $\alpha = 0.15$ in both the PageRank and the PPR algorithms.}.
Here, the former designates the number of opportunities in which each node can earn rewards as a curator for the state transition $\{G_{t}\}_{t=0}^{1000}$, while the latter designates the importance of each node from the viewpoint of the entire DAG in $G_{1000}$.
We specifically computed Spearman's rank correlation coefficient\footnote{We cannot use Pearson correlation coefficient because both frequency distribution and PageRank scores follow not normal distribution but power-law distribution.} of these values $10$ times\footnote{Correlation coefficients are different in each of $10$ computations because curators are assigned stochastically according to PPR algorithm, contrary to the constant PageRank score.} for each $20$ cases with a different number of assigned curators: $n = \{1,2, \cdots, 20\}$.

\begin{figure}[t]
    \centering
    \includegraphics[bb=0 0 3563 2581, width=0.5\hsize]{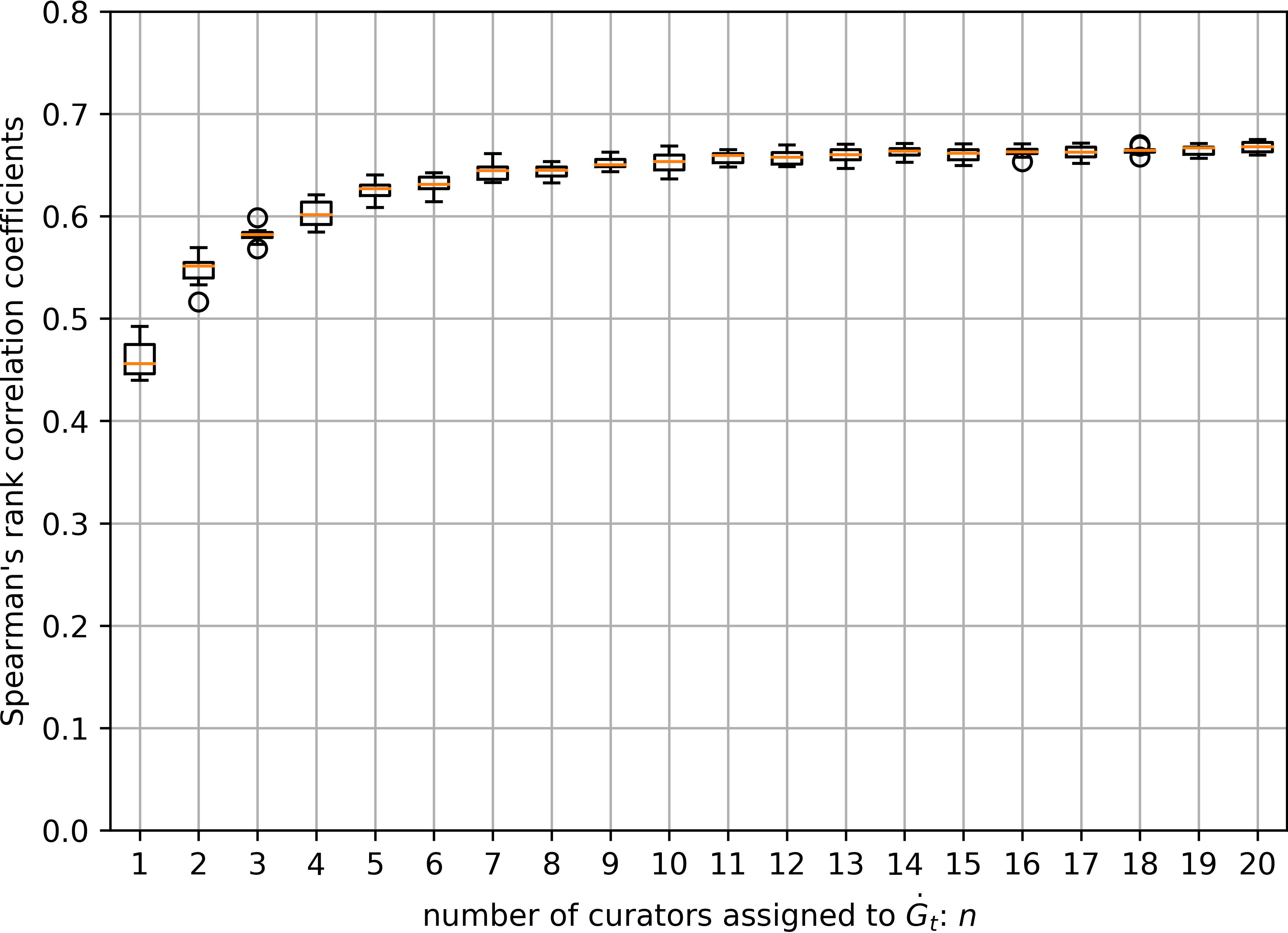}
    \caption{The first experiment computes Spearman's rank correlation coefficients between the frequency distribution of curator assignment up to $G_{1000}$ and the PageRank score to the DAG in $G_{1000}$. The box plot for all $200$ coefficients ($10$ times for each $n = \{1,2, \cdots, 20\}$) represents the moderate positive correlation, which increases as $n$ increases and converges between $0.65$ and $0.7$. This result supports our assumption that {\em CitedTCR} tends to select curators more frequently from nodes that are considered important in $G_{t}$.}
    \label{experiment1}
\end{figure}

Fig. \ref{experiment1} summarizes the trend of $200$ derived correlation coefficients in a box plot that depicts the median value as orange lines, $25/75$ percentile as boxes, pseudo-maximum/minimum value as bars, and outliers as circles.
This figure reveals that all correlation coefficients are within the range of $0.4$ to $0.7$, which can be regarded as moderately correlated. Moreover, they begin to converge between $0.65$ and $0.7$ when $n$ exceeds $10$.
These results indicates that {\em CitedTCR} can retain sufficient incentive to register high-quality content, especially when it assigns more than $10$ curators to ${\dot{G_{t}}}$, even though curator assignment relies on the PPR algorithm without base nodes.

\begin{figure}[t]
    \centering
    \includegraphics[bb=0 0 7854 5819, width=0.85\hsize]{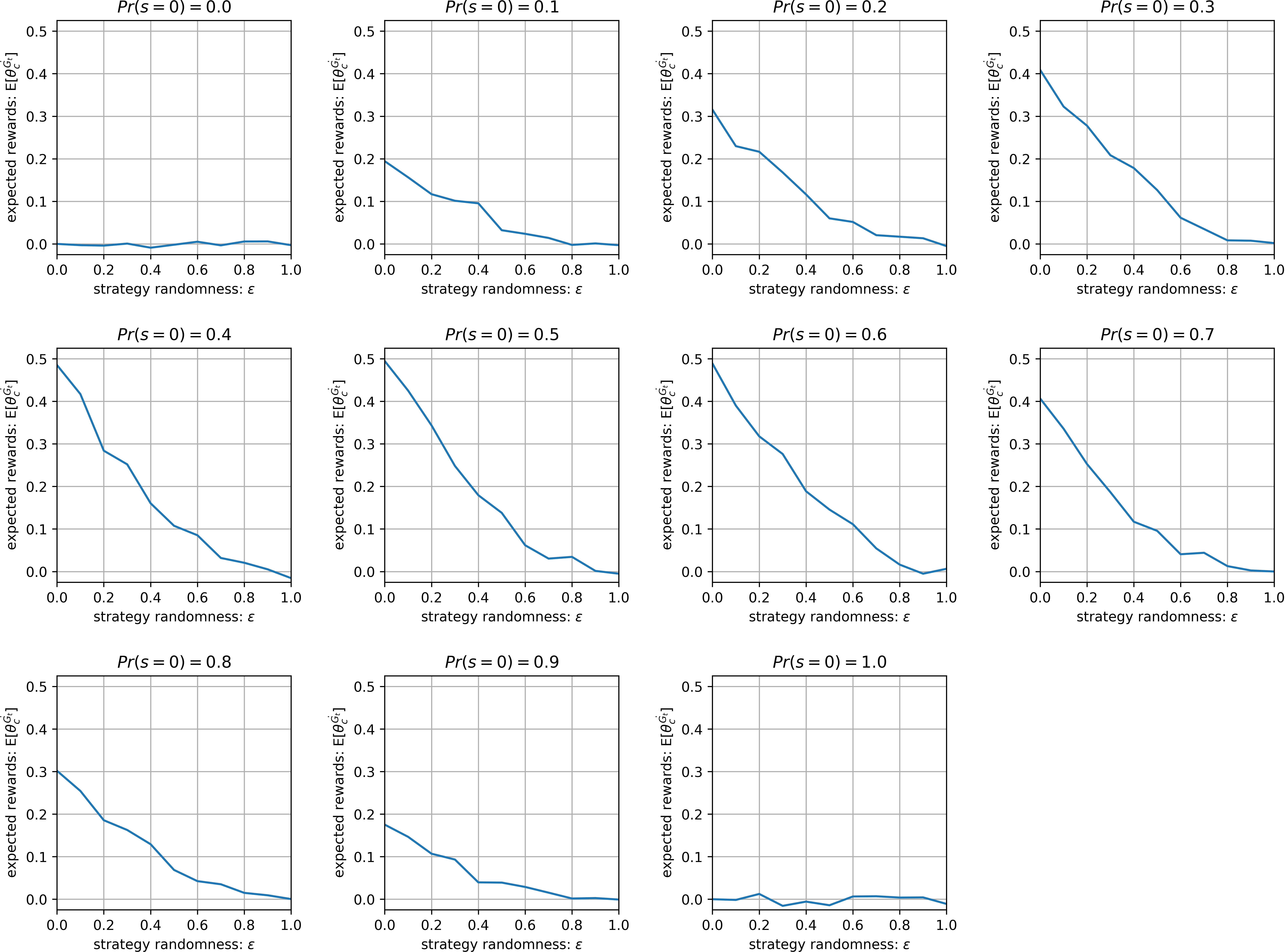}
    \caption{The second experiment computes $\mathbb{E}\left[ \theta_{c}^{\dot{G_{t}}} \right]$, which varies depending on strategy randomness $\epsilon$ and signal distribution $Pr(s=0)$ in $G_{t}$. $11$ graphs for $121$ $\{0.0,0.1, \cdots, 1.0\} \times \{0.0,0.1, \cdots, 1.0\}$ cases reveal that $\mathbb{E}\left[ \theta_{c}^{\dot{G_{t}}} \right]$ is maximized when all curators select the truthful strategy (i.e., $\epsilon = 0.0$), except the $Pr(s=0)=0.0$ or $1.0$ case. This result is consistent with {\em strong truthfulness}, which can elicit informative reports from assigned curators in {\em CitedTCR}.}
    \label{experiment2}
\end{figure}

\subsection{Incentive for eliciting informative reports}
After the simulation of curator assignment, the second experiment adds the DG13 mechanism to the first experiment to compute the expected reward $\mathbb{E}\left[ \theta_{c}^{\dot{G_{t}}} \right]$ stemming from $r_{c}^{\dot{G_{t}}}$ and $r_{\hat{c}}^{\dot{G_{t}}}$.
To simulate the settings of DG13, in which the user reports the received signal $s\in \{0,1\}$ according to a given strategy, we stochastically allocate the strategy and $s\in \{0,1\}$ in advance to all $1,421$ nodes.
In this experiment, the nodes are assumed to use either the truthful strategy or the aforementioned 50-50 uninformative strategy.
The allocation of the two strategies is subject to the exogenous randomness parameter $\epsilon = \{0.0, 0.1, \cdots, 1.0\}$, where the expected number of nodes with the uninformative strategy is $\epsilon \cdot 1,421$, and the expected number of nodes with the truthful strategy is $(1 - \epsilon)\cdot 1,421$.
Similarly, $s\in \{0,1\}$ is allocated to $1,421$ nodes by another exogenous parameter $Pr(s=0) = \{0.0,0.1, \cdots, 1.0\}$.
We computed $\mathbb{E}\left[ \theta_{c}^{\dot{G_{t}}} \right]$ by averaging the total reward generated in $\{G_{t}\}_{t=0}^{1000}$ for each of the $121$ environments comprising different allocations of these two exogenous parameters $\{0.0,0.1, \cdots, 1.0\} \times \{0.0,0.1, \cdots, 1.0\}$, in which $n = 10$ and $m = 0$ are fixed in any environment (i.e., ${\dot{G_{t}}}$ is always accepted into $G_{t}$ regardless of the reports).

Fig. \ref{experiment2} summarizes our results in $11$ graphs with different $Pr(s=0)$, which depicts the trend that $\mathbb{E}\left[ \theta_{c}^{\dot{G_{t}}} \right]$ in the same $Pr(s=0)$ is maximized when all curators use the truthful strategy (i.e., $\epsilon = 0.0$), even though the amount of maximized $\mathbb{E}\left[ \theta_{c}^{\dot{G_{t}}} \right]$ decreases as $Pr(s=0)$ deviates from $0.5$ and becomes indifferent with respect to $\epsilon$ if $Pr(s=0)=0.0$ or $1.0$\footnote{This is because truthful strategy becomes indifferent with an uninformative strategy that always returns $r = 0$ (or $1$), if $Pr(s=0)=0.0$ (or $1.0$) holds. Note that $Pr(s=0)=0.0$ and $1.0$ are outside the scope of DG13 as we cannot put the assumption of positively correlated signals on the environments.}.
This result is consistent with the {\em strong truthfulness} discussed in Section \ref{dg13explain} and indicates that {\em CitedTCR} retains incentive to elicit informative reports from assigned curators through DG13-based rewards.

\section{Conclusion} \label{conclusion}
In this study, we proposed {\em CitedTCR}, which incorporates the expertise of anonymous curators into existing TCRs by constructing a reliable citation graph, which is a common proxy for measuring the quality of technical content (e.g., academic papers, patents).
To achieve this enhancement on a public peer-to-peer network, we leveraged the PPR algorithm and DG13 mechanism, where the former assigns appropriate curators and the latter elicits informative reports from the assigned curators.
As a hybrid of network-based and token-based recommender systems, the combination of previous methods can lead to an incentive design that provides more reward tokens to users as they register high-quality content and continue producing informative reports.
Although this incentive design has a different approach than existing TCRs that involve token staking, {\em CitedTCR} has  sufficient utility, which was confirmed theoretically and experimentally.
This study can contribute to the emerging discussion on TCRs through its use of a citation graph and peer-prediction mechanism.

However, for practical implementation of this proposal, two remaining issues must be addressed in future work.
One involves relaxing the strong assumption of a one-to-one correspondence between users and nodes.
Despite the importance of being spam- and sybil-proof for the robustness of peer-to-peer systems, {\em CitedTCR} without one-to-one correspondence is vulnerable to such attacks because the role of the applicant and its curators can easily overlap if users can create many sybil accounts or post many contents to $G_{t}$ .
To overcome these attacks, an environment may be required in which curators are selected not from $V_{t}$, but from $U_{t}$, and $U_{t}$ has no incentive to create sybil accounts when posting multiple content.
The indices or algorithms for addressing similar issues have been proposed in the relevant fields of {\em CitedTCR} such as {\em SocialRank} \cite{tsai2014ranking} in network-based recommender systems, {\em h-index} \cite{hirsch2005index} in citation analysis, and {\em Proof of Stake} \cite{gui2018memo, saleh2018blockchain} in blockchain.
It is therefore a topic for future research to assess the availability of such existing studies in {\em CitedTCR}.

The second remaining task is to design a valuable reward token.
Although this study assumes that users act to maximize the amount of reward tokens, the power of tokens as an incentive is subject to their value, which is determined based on their utility, scarcity, and sustainability.
{\em CitedTCR} therefore requires additional mechanisms to ensure the value of reward tokens as in the Bitcoin protocol, where block-reward halving fixes total supply, and difficulty adjustment stabilizes hash rate.
A potential approach is to charge every applicant a token-based registration fee whose price is elastic and based on the frequency with which $\dot{G_{t}}$ is proposed in a given period\footnote{This concept corresponds to the difficulty adjustment in Bitcoin protocol.}.
This approach is worth considering as a registration fee gives the reward token a utility and can serve to prevent spam attacks.




\appendix
\setcounter{section}{0}
\section{Expected rewards in a simple token-staking scheme} \label{expected}
    Consider a simple token staking example in which $n$ curators stake a fixed $q$ number of tokens on one of the options.
    Let $k$ be the amount of (net) rewards that curators can obtain when their selections become the consensus, and let $p$ be the curators' subjective probability of the realization of this event. Then, the expected reward in this example is $\mathbb{E}(k) = pk - (1-p)q$.

    Specifically, $k$ is the redistribution of the total staked tokens $nq$ among the curators who have staked on the consensus with the exception of one's own stake, $q$.
    Accordingly, if we let $n^*$ be the number of curators who have staked on the consensus, $k = \frac{n}{n^*}q-q=\frac{n-n^*}{n^*}q$.
    By substituting this into the equation of $\mathbb{E}(k)$, we can derive the following condition:

    \begin{displaymath}
      \mathbb{E}(k) \left\{
        \begin{array}{c}
        > \\
        = \\
        <
        \end{array}
        \right\}
        0, \;\;{\rm if}\;\;
        \frac{p/(1-p)}{n^*/(n-n^*)} \left\{
        \begin{array}{c}
        > \\
        = \\
        <
        \end{array}
        \right\}
        1,
    \end{displaymath}

    \noindent
    where $\frac{p/(1-p)}{n^*/(n-n^*)}$ represents the odds ratio between the expected and actual value of the probability of one's choice becoming the consensus; i.e., the expected reward in the model takes a positive value only when we estimate the odds to be higher than their actual value and is zero as long as our estimation is precise (as a result of the zero-sum game).
    Furthermore, the expected reward under precise odds estimation is negative if we take the cost of curation into account\footnote{If we assume the cost of curation as $c$, the expected rewards in this example become $\mathbb{E}(k) = p(k-c) - (1-p)(q+c)$. This extension shifts the condition for $\mathbb{E}(k)=0$, from $\frac{p/(1-p)}{n^*/(n-n^*)}=1$ to $\frac{p/(1-p)}{n^*/(n-n^*)}=\frac{q+c}{q-\frac{n^*}{n-n^*}c}$, where the right-hand side of the new condition must be greater than one.}.

    These results reveal that the token-staking scheme does not have sufficient incentive to engage curators in consensus building.
    Providing new reward tokens to curators in proportion to the score of the peer-prediction mechanism is one possible approach to this problem.

\section{Proof of the strong truthfulness of the DG13 mechanism} \label{proof}
This proof uses notations that are compatible with Section \ref{dg13explain}.
The expected value of the reward term $\delta_{r_{c}^{\dot{G_{t}}},r_{\hat{c}}^{\dot{G_{t}}}}$ depends not only on the results of $r_{c}^{\dot{G_{t}}}$ and $r_{\hat{c}}^{\dot{G_{t}}}$, but also on the probability distribution of input signals that each node observes in period $t$, as follows:

\begin{displaymath}
    \mathbb{E}\left[\delta_{r_{c}^{\dot{G_{t}}},r_{\hat{c}}^{\dot{G_{t}}}}\right] = \sum_{s_{c}=0}^{1} \sum_{s_{\hat{c}}=0}^{1} Pr(s_{c},s_{\hat{c}})\cdot \delta_{r_{c}(s_{c}),r_{\hat{c}}(s_{\hat{c}})},
 \end{displaymath}
where $Pr(s_{c},s_{\hat{c}})$ is the joint probability distribution of the signals that $c$ and $\hat{c}$ can receive from $\dot{G_{t}}$.
Note that the right-hand side does not require superscript $\dot{G_{t}}$ because of the assumption of positively correlated signals.

As described in Section \ref{dg13explain}, the penalty term is the result of the comparison between two randomly picked reports that $c$ and $\hat{c}$ produce prior to period $t$.
We can write the expected value of the penalty in a similar form to the reward term as follows:
\begin{displaymath}
    \mathbb{E}\left[\delta_{r_{c}\in \{R_{c}\setminus \dot{R_{t}}\},r_{\hat{c}}\in \{R_{\hat{c}}\setminus \dot{R_{t}}\}}\right] = \sum_{s_{c}=0}^{1} \sum_{s_{\hat{c}}=0}^{1} Pr(s_{c})Pr(s_{\hat{c}})\cdot \delta_{r_{c}(s_{c}),r_{\hat{c}}(s_{\hat{c}})}.
\end{displaymath}
This uses product distribution $Pr(s_{c})Pr(s_{\hat{c}})$ rather than joint distribution $Pr(s_{c},s_{\hat{c}})$ because the penalty term covers all intertemporal reports included in $R_{c}\setminus \dot{R_{t}}$ and $R_{\hat{c}}\setminus \dot{R_{t}}$.

Consequently, $\mathbb{E}( \theta_{c}^{\dot{G_{t}}})$ can be expressed as
\begin{displaymath}
    \mathbb{E}\left[ \theta_{c}^{\dot{G_{t}}} \right] = \sum_{s_{c}=0}^{1} \sum_{s_{\hat{c}}=0}^{1} \left[Pr(s_{c},s_{\hat{c}}) - Pr(s_{c})Pr(s_{\hat{c}})\right]\cdot \delta_{r_{c}(s_{c}),r_{\hat{c}}(s_{\hat{c}})}.
\end{displaymath}
The terms in square brackets correspond to the correlation of $s_{c}$ and $s_{\hat{c}}$.
If one assumes that $Pr(s_{c},s_{\hat{c}}) - Pr(s_{c})Pr(s_{\hat{c}}) > 0$, then both $Pr(s_{c}|s_{\hat{c}}) > Pr(s_{c})$ and $Pr(s_{\hat{c}}|s_{c}) > Pr(s_{\hat{c}})$ hold because $Pr(s_{c},s_{\hat{c}}) = Pr(s_{c}|s_{\hat{c}})Pr(s_{\hat{c}}) = Pr(s_{\hat{c}}|s_{c})Pr(s_{c})$, i.e., $s_{c}$ and $s_{\hat{c}}$ are positively correlated in this case.

Because DG13 assumes positively correlated binary signals, the following condition holds in the expanded form of $\mathbb{E}( \theta_{c}^{\dot{G_{t}}})$:
\begin{align}
    \mathbb{E}\left[ \theta_{c}^{\dot{G_{t}}}\right] &= \left[Pr(s_{c}=0,s_{\hat{c}}=0) - Pr(s_{c}=0)Pr(s_{\hat{c}}=0)\right]_{>0}\cdot \delta_{r_{c}(0),r_{\hat{c}}(0)}\nonumber \\
    &+ \left[Pr(s_{c}=0,s_{\hat{c}}=1) - Pr(s_{c}=0)Pr(s_{\hat{c}}=1)\right]_{<0}\cdot \delta_{r_{c}(0),r_{\hat{c}}(1)}\nonumber \\
    &+ \left[Pr(s_{c}=1,s_{\hat{c}}=0) - Pr(s_{c}=1)Pr(s_{\hat{c}}=0)\right]_{<0}\cdot \delta_{r_{c}(1),r_{\hat{c}}(0)}\nonumber \\
    &+ \left[Pr(s_{c}=1,s_{\hat{c}}=1) - Pr(s_{c}=1)Pr(s_{\hat{c}}=1)\right]_{>0}\cdot \delta_{r_{c}(1),r_{\hat{c}}(1)},\nonumber
\end{align}
where $\left[x\right]_{>0}$ and $\left[x\right]_{<0}$ indicate that $x$ is positive and negative, respectively\footnote{Furthermore, if we designate $Pr(s_{c}=0,s_{\hat{c}}=0)-Pr(s_{c}=0)Pr(s_{\hat{c}}=0)=P_{00}, Pr(s_{c}=0,s_{\hat{c}}=1)-Pr(s_{c}=0)Pr(s_{\hat{c}}=1)=P_{01},Pr(s_{c}=1,s_{\hat{c}}=0)-Pr(s_{c}=1)Pr(s_{\hat{c}}=0)=P_{10}, Pr(s_{c}=1,s_{\hat{c}}=1)-Pr(s_{c}=1)Pr(s_{\hat{c}}=1)=P_{11}$, they have the following relations: $P_{00}=P_{11}, P_{01}=P_{10}, P_{00}+P_{01}+P_{10}+P_{11}=0$.}.

It is apparent that $\mathbb{E}( \theta_{c}^{\dot{G_{t}}})$ is maximized only when both $c$ and $\hat{c}$ provide truthful reports ($r(0)=0, r(1)=1$) or opposite reports ($r(0)=1, r(1)=0$).
Any other pattern, such as nodes using asymmetric strategies or always reporting the same signal, produces less expected values.
Under the assumption of using one reporting strategy, this outcome indicates that $\mathbb{E}( \theta_{c}^{\dot{G_{t}}})$ is maximized only when both $x$ and $\hat{c}$ adopt either a truthful or opposite strategy.
Thus, DG13 satisfies strong truthfulness. \hfill$\qed$

\end{document}